\documentclass[aps,prl,twocolumn]{revtex4}
\usepackage{graphicx}

\begin{document}
\title{Entropically-stabilised growth of a two-dimensional random tiling}
\author{Andrew Stannard, Matthew O. Blunt, Peter H. Beton, and Juan P. Garrahan}
\affiliation{School of Physics and Astronomy, University of Nottingham, University Park, Nottingham, NG7 2RD UK}

\begin{abstract}
The assembly of molecular networks into structures such as random tilings and glasses has recently been demonstrated for a number of two-dimensional systems. These structures are dynamically-arrested on experimental timescales so the critical regime in their formation is that of initial growth.  Here we identify a transition from energetic to entropic stabilisation in the nucleation and growth of a molecular rhombus tiling.  Calculations based on a lattice gas model show that clustering of topological defects and the formation of faceted boundaries followed by a slow relaxation to equilibrium occurs under conditions of energetic stabilisation. We also identify an entropically-stabilised regime in which the system grows directly into an equilibrium configuration without the need for further relaxation. Our results provide a methodology for identifying equilibrium and non-equilibrium randomness in the growth of molecular tilings, and we demonstrate that equilibrium spatial statistics are compatible with exponentially slow dynamical behaviour. 
\end{abstract}


\maketitle

The properties of two-dimensional supramolecular networks have been the focus of growing interest in recent years with most efforts directed towards the controlled introduction of translational order into such systems \cite{Elemans2009, Bartels2010}. However there have been several recent observations of surface-bound supramolecular arrays which assemble into dynamically-arrested structures akin to glasses \cite{Blunt2008, Otero2008, Marschall2010} which lack translational order. Such arrangements raise many interesting questions related to the growth of random systems \cite{Swallen2007, Bindi2009}. In particular it is important to distinguish randomising effects which arise from kinetic effects, such as nucleation \cite{Hwang1997, Brune1998}, from equilibrium disorder due to entropic terms in the free energy. Entropically-stabilised disorder may be regarded as intrinsic randomness, whereas kinetically-driven disorder is often determined by sample history and preparative conditions. In one recent study \cite{Blunt2008} a random molecular rhombus tiling was shown to have equilibrium (maximum entropy) spatial correlations, despite being frozen on an experimental timescale. In such a system the maximum entropy configuration must form, and be frozen in, during the initial growth, since the spatio-temporal fluctuations which normally facilitate the evolution of kinetically-trapped configurations to equilibrium are absent. However, it is not clear, \textit{a priori}, that there is a set of local rules for molecular attachment which can lead to the direct growth of a `perfect', i.e. maximum entropy, configuration.

In this paper we address this question and show that equilibrium and non-equilibrium effects in the growth of a rhombus tiling \cite{Fisher1961, Kasteleyn1963, Blote1982, Henley1999, Fisher1963, Destainville1998, Cohn2001, Jacobsen2009} may be distinguished using tile-tile correlations of arrays simulated using a lattice gas model \cite{Garrahan2009}. Direct growth to a configuration with equilibrium statistics occurs when entropic terms dominate the free energy, while non-equilibrium effects result in faceted islands and clustering of topological defects.

The parameters which control growth are the tile-tile interaction energy, $\varepsilon$, the tile adsorption energy, $\mu$, and the temperature, $T$. We consider (see Fig.~1) a triangular lattice with sites, labelled $i$, that are either occupied by half a tile or vacant. Each rhombus tile occupies two adjacent sites and lies in one of three orientations (distinguished by different colours). For molecules deposited from solution \cite{Blunt2008} $\mu$ corresponds to the difference between the net adsorption and solvation energies of the molecule ($\mu>0$ implies a preference for solvation). The binding energy per tile for a completely tiled surface is $E_{bind}=-(2\varepsilon -\mu)$.

\begin{figure}
\includegraphics[width=8cm]{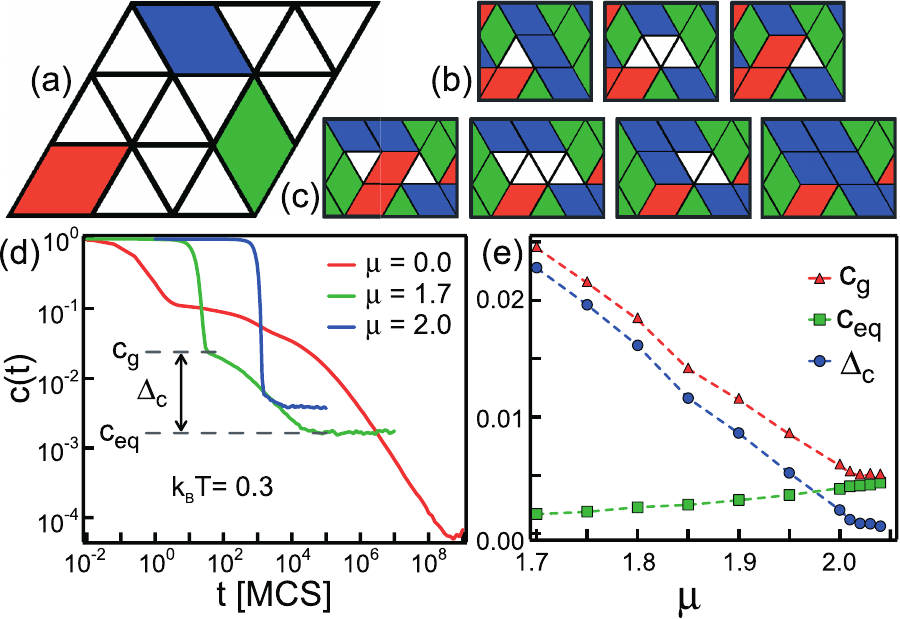}
\caption{Schematic of: (a) lattice and tiles; (b) defect diffusion mediated by tile detachment and re-attachment; (c) annihilation of a defect pair (left to right) or generation of defect pair (right to left). (d) The fraction of empty sites as a function of time (in units of Monte Carlo sweeps) ($k_BT=0.3$, lattice size $N=10^6$. (e) Dependence of $c_g$, $c_{eq}$, and $\Delta c$ [defined in text and in (d)] on $\mu$.}
\end{figure} 

Using a Metropolis algorithm \cite{Newman1999}, sites are chosen randomly and, if empty, a tile is added with probability $e^{-\Delta E/k_BT}$ for $\Delta E>0$ and with unit probability for $\Delta E \leq 0$, where $\Delta E$ is the associated change in energy. If the site is occupied, tile removal is accepted with probability $e^{-\Delta E/k_BT}/3$. The factor of $1/3$ ensures detailed balance is satisfied. The energy required to remove a tile with $p$ nearest neighbours is $E_{rem}=(p\varepsilon -\mu)$, which, for $\mu>0$, may be either positive or negative depending on the local environment. One Monte Carlo sweep (MCS) corresponds to the random inspection of $3N$ sites of the (rhomboid) lattice ($N$ is the maximum number of adsorbed tiles), and sets the unit of time. Periodic boundary conditions are used and all energies are henceforth expressed in units of $\varepsilon$. This is a generalisation, through the introduction of the parameter, $\mu$, of a model previously used to show that rhombus tilings are glassy \cite{Garrahan2009}.

As the time increases, the fraction of empty lattice sites, $c(t)$, reduces from 1 (empty lattice) and eventually relaxes to an equilibrium phase at a $\mu$-dependent constant value of $c(t)$, $c_{eq}$.  In the initial growth phase $c(t)$ falls until an abrupt change in gradient occurs. At this point, to a good approximation, there are no more available vacancies (neighbouring pairs of unfilled triangles) which could directly accommodate a tile. However, the lattice is not completely tiled and triangular void defects are also present. These are topological defects with two effective charges corresponding to triangles pointing up and pointing down \cite{Fisher1963, Linde2001, Krauth2003}. Further relaxation is mediated by defect diffusion and annihilation (neighbouring defects of opposite effective charge form a vacancy which may be occupied by a tile - see Fig.~1) \cite{Garrahan2009}. Since defect diffusion is an activated process (barrier $3-\mu$) there is a slowing down which gives rise to the clear change in gradient discussed above. In the equilibrium regime there is a dynamic balance between the generation of triangular defect pairs (from the removal of tiles) and their diffusion and annihilation \cite{Garrahan2009}.

We parameterise the value of $c(t)$ after the initial growth phase as $c_g$ (see Fig.~1 where the values for $c_g$ and $c_{eq}$ are identified for the $\mu=1.7$ curve). We also introduce a parameter, $\Delta c(=c_g-c_{eq})$, to quantify the difference between the defect density in equilibrium and immediately after the initial growth phase. The dependence of the parameters $c_g$, $c_{eq}$, and $\Delta c$ on $\mu$ is shown in Fig.~1(e) over the parameter range $1.7<\mu<2.1$. As expected, $c_{eq}$ increases with increasing $\mu$ since the energy barrier for tile detachment is reduced. Interestingly, in the range $\mu>2$ the binding energy, $E_{bind}>0$, and no tiling would be expected for an ordered system. However, random tilings do grow in this regime; the variation of $c(t)$ for $\mu=2$ is shown in Fig.~1(d), and values for $c_g$, $c_{eq}$ and $\Delta c$ extracted in the regime where $E_{bind}>0$ ($\sim 2.1>\mu>2$), are shown in Fig.~1(e). We show below that in this regime entropic contributions lead to a free energy, $F$, given by $(NE_{bind}-TS)$ where $S$ is the entropy, which can be negative, favouring a tiling, even when $E_{bind}>0$ (Joseph \textit{et al.} \cite{Joseph1997} make similar arguments in the context of entropically-stabilised quasicrystals \cite{Henley1999, Widom1989}). Furthermore, in this entropically-stabilised regime, $\Delta c\rightarrow 0$, implying, as confirmed below, that the initial growth phase leads directly to an equilibrium regime.

We now consider the differences in nucleation, morphology and tile statistics in the energetically-stabilised ($\mu<2$) and entropically-stabilised ($\mu>2$) regimes, focussing first on low values of $\mu$. In Fig.~2 we show islands which have been nucleated and are growing in the initial growth regime. For $0<\mu<1$, $E_{rem}$, the barrier for tile removal, is positive even for $p=1$ indicating that any nucleated island formed by two neighbouring tiles is stable (note for $\mu \leq 0$ even isolated tiles are stable nuclei). Accordingly, the simulated growth in this regime [Fig.~2(a)] shows a large number of small, irregular islands of tiles.  This is an essentially homogeneous growth regime: very quickly the islands merge forming an imperfect tiling of the plane.

\begin{figure}
\includegraphics[width=8cm]{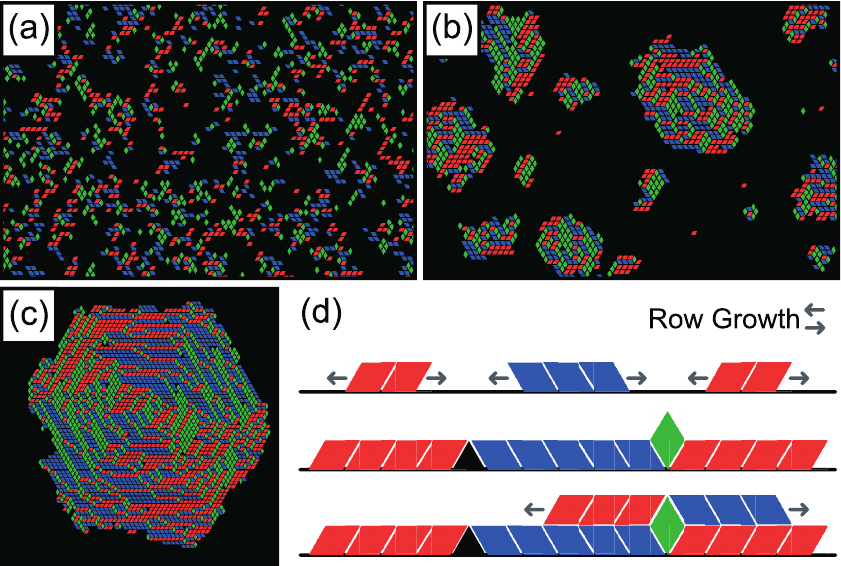}
\caption{(a-c) Simulated growth of tilings for varying $\mu$ with $k_BT=0.2$: (a) $\mu=0.5$, high nucleation density, irregular islands ($c(t)=0.7$); (b) $\mu=1.5$, reduced nucleation density, facetted islands ($c(t)=0.7$); (c) $\mu=1.8$, example of an inhomogeneous tile distribution within strongly facetted island; (d) schematic of growth along a straight interface for $1<\mu<2$.}
\end{figure}

For $1<\mu<2$ we have heterogeneous growth; as $\mu$ increases, islands become larger and their number decreases.  Furthermore, the islands that form are faceted and hexagonal [see Fig.~2(b)] with a clear deficit of tiles in one of the three possible orientations (colours) in each of the six triangular segments of the island [see Fig.~2(c)]. The faceting occurs since the smallest energetically stable nucleus requires a minimum of three tiles in a hexagonal configuration. Outward growth results in the (approximate) propagation of the hexagonal shape since growth along an edge favours the addition of a row formed by one of the two tile orientations with an edge parallel to the island boundary. For example, in Fig.~2(d), rows of blue and/or red tiles grow and where they meet an upward- or downward-pointing triangular defect is formed. The downward defect is trapped, but the upward defect may be occupied by a green tile, leading to an excess of defects of one effective charge (downward-pointing in this case) in each segment. Inspection of the hexagonal island in Fig.~2(c) shows rows of tiles of two colours with either defects, or a tile of the third colour where they meet, consistent with this simple explanation. Thus the faceting is due to minimisation of the boundary energy, but also results in clustering of defects with the same effective charge. A local imbalance of tiles gives rise to an increase in entropic free energy, and is not expected for an equilibrium configuration. We note the interesting similarity between the growth of energetically-favoured hexagonal tilings observed here with the `arctic circle' problem in rhombus tilings subject to hexagonal confining boundaries \cite{Destainville1998, Cohn2001}.

As growth in this regime continues the islands merge while maintaining, approximately, the primordial structure introduced in the nucleation stage. An example is shown in Fig.~3(a) which shows the tiling that is formed immediately after all the growing islands merge.  Defects in this tiling are not distributed uniformly as in equilibrium \cite{Jacobsen2009}.  Fig.~3(b) shows the topological charge density for the tiling of Fig.~3(a).  Defect clustering is evident in the large variations in topological charge.  The original nucleation sites within the tiling can be identified as singular points in this defect density [Fig.~3(b)], which confirms the spatial correlation of defect clustering and nucleation sites. Further temporal evolution governed by defect diffusion and annihilation \cite{Garrahan2009} leads to equilibration of the tiling, see Fig.~3(c), when $c(t)$ reaches its equilibrium value [see Fig.~3(d)].

\begin{figure}
\includegraphics[width=8cm]{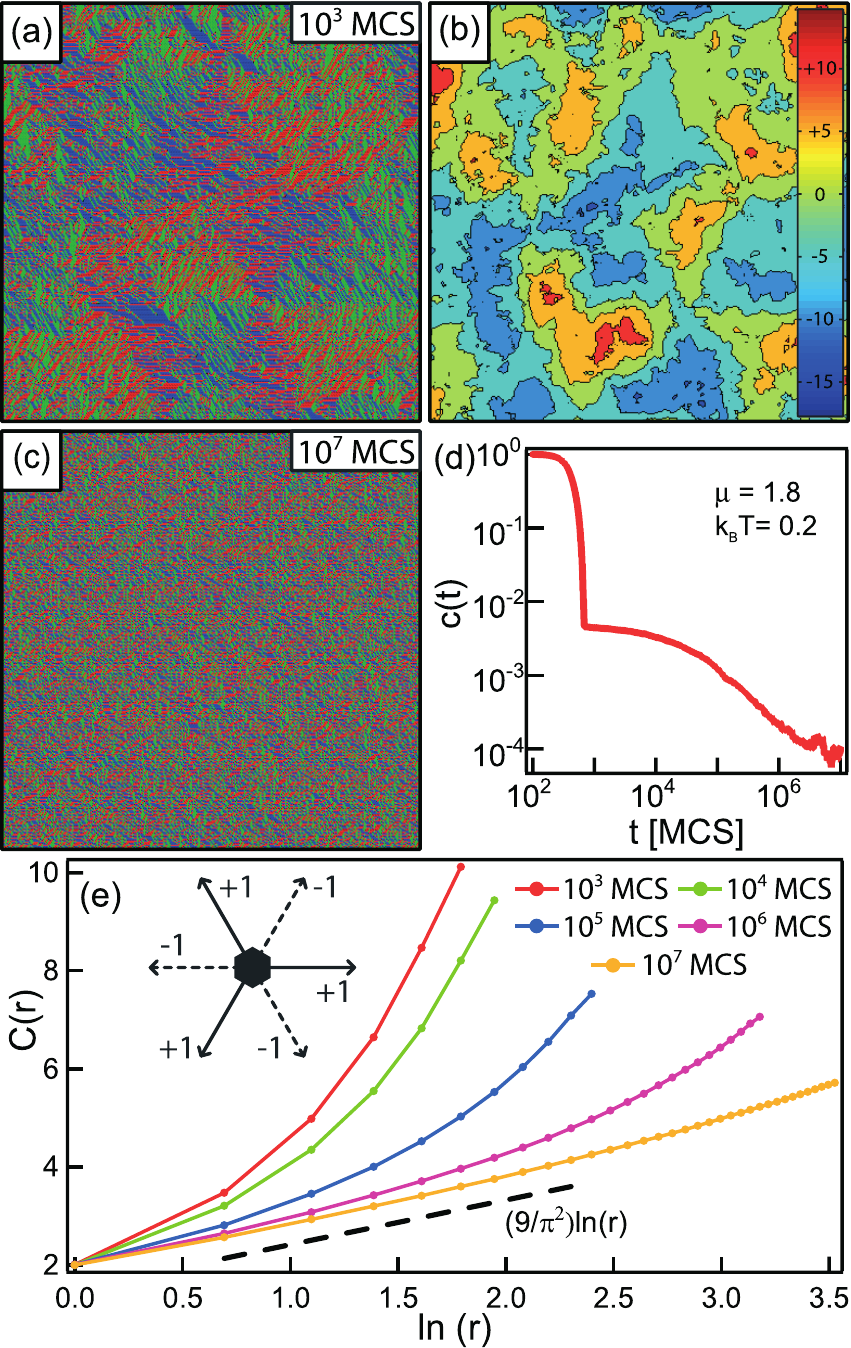}
\caption{(a) An inhomogeneous tiling resulting from strongly faceted growth of multiple islands, $k_BT=0.2$, $\mu=1.8$, $N=1.6\times10^5$ ($\sim4\times10^4$ tiles shown). (b) Clustering of topological charge shown in a charge density map. The value at a point corresponds to the number of upward-pointing minus downward-pointing defects within a range of three times the average defect separation. (c) The tiling after relaxing to an entropically-maximized equilibrium state ($\sim4\times10^4$ tiles shown), (d) the corresponding $c(t)$ behaviour, and (e) height correlation functions during relaxation, indicating convergence to maximum randomness.}
\end{figure}

The tilings are analysed using a lifting dimension \cite{Henley1999, HenleyPreprint} in which an effective height, $h(\vec{r})$ is assigned to each vertex (with in-plane co-ordinates, $\vec{r}$) in the tiling.  The height is calculated using the scheme shown in the Fig.~3(e) inset in which a displacement along a rhombus edge leads to a change in height of $\pm 1$. The height correlation function, $C(r)=\langle [h(0)-h(r)]^2 \rangle$, can be calculated and for a maximally random tiling, $C(r)=(\pi K_0)^{-1} \ln(r)+c$, has a logarithmic dependence on position, where $c$ is a constant and $K_0=\pi/9$ \cite{Henley1999}. Fig.~3(e) shows the correlation functions during the simulated growth of the tilings in Figs.~3(a)~\&~3(c). For the tiling in Fig.~3(a) the correlation function is not logarithmic. For increasing times the correlation functions approach a linear dependence on ln(r), with the expected gradient $9/\pi^2$, confirming that the final configuration [Fig.~3(c)] is equilibrated. This supports the hypothesis that a logarithmic dependence is associated with an equilibrium configuration rather than kinetically-controlled randomness, but the exponentially slow approach to equilibrium cannot account for tilings which are both dynamically-arrested and maximum-entropy.

For $\mu \geq 2$ nucleated islands do not show faceting or inhomogeneities [Fig.~4(a) inset] and our simulations show direct growth into a maximum-entropy configuration. In Fig.~4(a) we plot correlation functions for tilings immediately after the initial growth regime is completed [determined by the change in gradient in $c(t)$] and find an approach to a logarithmic dependence on $r$ as $\mu$ increases. Note from Fig.~1(e) that $\Delta c \rightarrow 0$ for $\mu>2$ and these results confirm that this simple parameter provides a reliable indicator for a regime of direct growth into an equilibrium configuration without the requirement for defect-mediated relaxation.

\begin{figure}
\includegraphics[width=8cm]{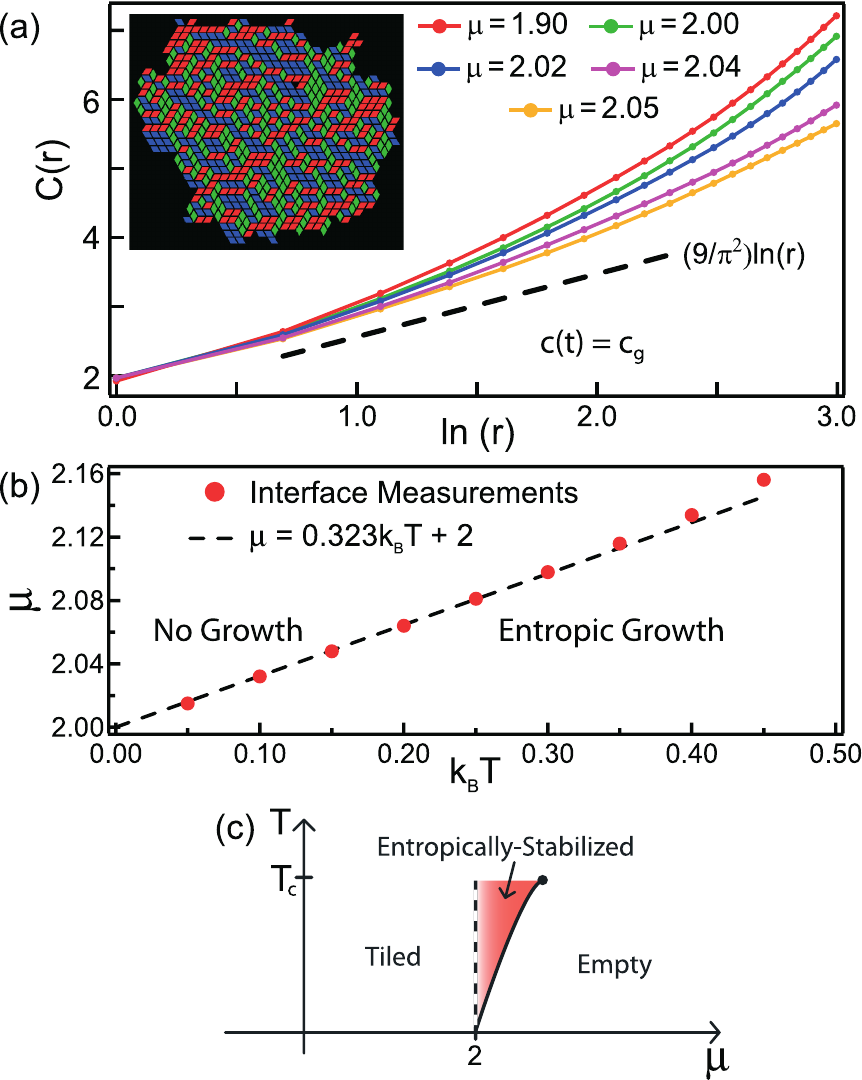}
\caption{(a) Height correlation functions calculated for tilings immediately after the initial growth stage [where $c(t)=c_g$] at $k_BT=0.3$ with varying $\mu$. (b) Calculations of stationary interfaces marking the tiled-untiled phase equilibrium. (c) Illustrative representation of the $\mu-T$ parameter space, indicating the region where tilings are entropically-stabilized.}
\end{figure}

To confirm that tilings for $\mu>2$ are entropically stabilised we need to establish the equilibrium phase boundary $\mu(T)$ between the tiled and empty phases.  We establish this by investigating whether an interface between an equilibrated tiling and an empty lattice recedes (no growth) or propagates (growth) \cite{Newman1999}. The value of $\mu$ where this transition occurs is plotted against temperature in Fig.~4(b). As discussed above, for an ordered system no growth is expected for $\mu>2$, but the free energy may be negative when $E_{bind}>0$, if $S_{tile}>E_{bind}/T$, where $S_{tile}$ is the entropy per tile, or $\mu<2+TS_{tile}$. Our simulations [Fig.~4(b)] give $S_{tile}=0.32k_B$ in excellent agreement with the ideal value for the rhombus tiling entropy density of  $0.323066k_B$ \cite{Henley1999, Wannier1950}.

These results are highly relevant to recent experiments \cite{Blunt2008}. We propose that, in both the energetically- and entropically-stabilised regimes, growth proceeds through the initial regime to the point identified in Fig.~1 where there is a slowing down of the evolution of the tiling. Further evolution is determined by the barrier to defect propagation, ($3-\mu$). If this is small compared with $k_BT$, equilibration can occur through defect propagation. However, for many molecular systems the barrier is at least an order of magnitude greater than the thermal energy and the configuration is therefore dynamically-arrested with spatial statistics which are frozen immediately after the initial tiling of the surface. A broad range of possible spatial distributions can occur, including, for $\mu\sim2$, the recently-observed maximum-entropy arrangement \cite{Blunt2008}. However for lower values of $\mu$ a configuration with a frozen-in, non-equilibrium spatial distribution of tiles [comparable to Fig.~3(a)] might be attainable in experiments.

The phase behaviour observed in these simulations is summarised in Fig.~4(c) and invites analogy with magnetic Ising systems since the total energy of a partially tiled surface is given by $E=-\frac{\varepsilon}{2}\sum_{i=1}^{2N}n_i\left(\sum_{j=1}^3n_j-1\right)+\frac{\mu}{2}\sum_{i=1}^{2N}n_i$, where $n_i=1(0)$ for an occupied (unoccupied) site (for a related example of the application of the lattice gas model to adsorbed molecular layers see Ref. \cite{Tao2008}). The index $i$ runs over all triangular sites and $j$ runs over the three nearest neighbours of site $i$. The tile-tile interaction, $\varepsilon$, is analogous to the spin-spin coupling, normally denoted by $J$, $\mu$ is analogous to magnetic field, and $n_i$ to spin state. In Fig.~4(c) a boundary at $\mu=2$ shows the threshold above which the internal energy is positive. However for $T<T_c\sim\varepsilon/k_B$ there is an entropically-stabilised regime for the random rhombus tiling [shaded region in Fig.~4(c)]. This phase boundary can be determined in our simulations up to $k_BT\sim0.5$ which we identify as an approximate critical temperature for this transition. The deposition of a molecular layer is thus equivalent to a quench from high $\mu$ (analogue magnetic field).  The growth dynamics after such a quench allows an investigation of this phase diagram for systems where dynamics are slow.

Our results show that equilibrium and non-equilibrium randomisation may be distinguished for the rhombus tiling. Moreover we have shown that equilibrium spatial statistics may occur even for dynamically-arrested systems although other outcomes such as faceting and defect clustering are also possible. These results have general relevance for molecular layers adsorbed at a liquid-solid interface where it has previously been assumed, correctly in many cases, that a dynamic equilibrium is established with molecules continually exchanged between solvated and adsorbed states (see \cite{deFeyter2003} for example). Our results show that such exchange is not required for the formation of maximum-entropy arrangements. There are also interesting links between entropically-stabilised growth and several other problems in bio- and condensed matter physics, such as the crystallisation of anisotropic particles \cite{WhitelamPreprint}.   Furthermore, molecular rhombus tiles provide a new system to explore, both experimentally and theoretically, equilibrium and non-equilibrium behaviour in connection with `Coulomb' and other exotic phases which can exhibit fractional excitations \cite{HenleyPreprint}, such as frustrated magnets with effective magnetic monopoles \cite{Castelnovo2008}, quasi-crystals \cite{Bindi2009}, and glasses \cite{Swallen2007, Chandler2010}. 

We thank the EPSRC for financial support through grants EP/D048761/1 and EP/P502632/1.

\end{document}